\documentclass[conference]{IEEEtran}

\IEEEoverridecommandlockouts

\usepackage[utf8]{inputenc}
\usepackage[T1]{fontenc}

\usepackage{amsmath}
\usepackage[ruled,vlined,linesnumbered]{algorithm2e}
\SetKwFor{ForEach}{for each}{do}{end}
\usepackage{graphicx}
\usepackage{textcomp}
\usepackage{xcolor}
\usepackage{booktabs}
\usepackage{multirow}
\usepackage{tabularx}
\usepackage{array}
\usepackage{url}
\usepackage{cite}
\usepackage{hyperref}

\def\BibTeX{{\rm B\kern-.05em{\sc i\kern-.025em b}\kern-.08em
    T\kern-.1667em\lower.7ex\hbox{E}\kern-.125emX}}

\begin{document}

    \title{SSR: Structural Signal Retriever for Scalable Agent Discovery in LLM-Orchestrated Multi-Agent Systems}
    
    \author{
            \IEEEauthorblockN{Giancarlo Maldonado Cárdenas, Carlos A. Astudillo, and Nelson L. S. da Fonseca}
            \IEEEauthorblockA{Institute of Computing -- State University of Campinas, Brazil}
            \IEEEauthorblockA{E-mails: giancarlo@lrc.ic.unicamp.br, castudillo@unicamp.br, and nfonseca@ic.unicamp.br}
        }
    
    \maketitle
    
\begin{abstract}
Large-scale multi-agent ecosystems require mechanisms for discovering suitable agents without exposing entire registries to a large language model (LLM) orchestrator. The challenge is not only to retrieve relevant Agent Cards from a growing registry, but also to maintain a compact candidate context, since each additional candidate increases prompt size, inference cost, latency, and the risk of exceeding the model context window. We formulate this problem as \emph{candidate-context
scalability} and propose the \emph{Structural Signal Retriever} (SSR), our deterministic and training-free retrieval method designed for Agent Card discovery. SSR derives structural lexical signals from capability-bearing Agent Card fields, weights them using inverse document frequency (IDF), and uses an Inverted Signal Index (ISI) for selective candidate scoring. We evaluate SSR using a synthetic agent benchmark, a public MS MARCO adaptation, and end-to-end LLM orchestration. At 1,000 agents, SSR reduces LLM input by 98.8\% and orchestration latency by 70.9\% relative to full-catalog orchestration.
On MS MARCO, IDF improves retrieval effectiveness within the
structural representation, while the ISI preserves retrieval
effectiveness and reduces explicitly scored passages by 96.8\%. End-to-end experiments show that candidate
pre-selection maintains a compact candidate context as registry size grows and avoids the operational failures observed with full-catalog orchestration. These results support SSR as an Agent-Card-native, traceable retrieval mechanism for candidate-context scalability.
\end{abstract}

 
\section{Introduction}

Recent advances in large language models (LLMs), interoperable agent
protocols, and machine-readable capability descriptions are enabling
multi-agent ecosystems in which autonomous agents can be dynamically
discovered and orchestrated
\cite{wang2025ioaSurvey,chen2025ioa}. Protocols such as the Model
Context Protocol (MCP) and Agent2Agent (A2A) expose structured
descriptions of tools and agents \cite{mcp2025,a2a2025}. In particular,
A2A defines an \emph{Agent Card} as a machine-readable description of
an agent's identity, capabilities, skills, interfaces, and related
metadata \cite{a2a2025}. We use \emph{Agent Card} throughout this paper
to denote such structured capability descriptions.

A straightforward orchestration strategy is to present every available
Agent Card to an LLM and ask it to select the most suitable agent.
However, this full-catalog approach directly couples registry size to
the amount of information presented to the LLM orchestrator. As the
registry grows, more Agent Cards must be incorporated into the prompt,
increasing context consumption, inference cost, and latency, while
eventually reaching practical model or API limits. Related
context-scaling concerns have motivated retrieval and candidate
filtering in large tool collections
\cite{shi2025toolret,molfetta2025ports,liu2026toolscope}.

Candidate pre-selection itself is therefore not new. Existing work has
explored large-scale tool retrieval, context-aware filtering, learned
routing, and relational tool--agent retrieval
\cite{shi2025toolret,liu2026toolscope,lumer2025tooltoagent,
nizar2026agentasagraph,yao2026acerouter}. Our focus is complementary:
scalable agent discovery must account not only for retrieving relevant
Agent Cards from a growing registry, but also for controlling the
amount of candidate information ultimately presented to the LLM
orchestrator. We formulate this systems problem as
\emph{candidate-context scalability}: registry-scale retrieval should
remain operationally independent of the compact candidate context
subsequently presented to the LLM. Under this formulation, retrieval
may operate over the complete registry, while a maximum candidate
budget $K$ controls the number of top-ranked Agent Cards forwarded to
the LLM orchestrator. This separation makes retrieval coverage,
retrieval computation, and orchestration context distinct and
independently measurable components of the discovery pipeline.

To address this problem, we propose the \emph{Structural Signal
Retriever} (SSR), our deterministic and training-free retrieval method
designed specifically for Agent Card discovery. SSR constructs an
explicit retrieval representation directly from capability-bearing
fields already present in Agent Cards. It deterministically derives
structural lexical signals from these fields, weights matched signals
using inverse document frequency (IDF), and uses an Inverted Signal
Index (ISI) to restrict scoring to Agent Cards that share structural
evidence with the user query. The resulting candidates are ranked and
a compact pool of up to $K$ top-ranked Agent Cards is presented to the
LLM orchestrator for final agent selection. This design decouples the
growth of the agent registry from the amount of candidate information
processed by the LLM while preserving a direct correspondence between
each retrieval score and the explicit Agent Card signals that produced
it.

The methodological contribution of SSR does not lie in IDF weighting, inverted indexing, or candidate pre-selection individually. Rather, SSR integrates these established retrieval mechanisms into a dedicated pre-orchestration retrieval layer for Agent Card discovery in LLM-based multi-agent infrastructures. Within this layer, capability-bearing Agent Card fields provide explicit structural signals for candidate generation and ranking, while a maximum candidate budget controls the compact candidate context subsequently presented to
the LLM orchestrator.

To distinguish improvements arising from candidate pre-selection in
general from properties specific to SSR, the end-to-end evaluation
also instantiates the same candidate-selection architecture using the Best Matching 25 (BM25) ranking function \cite{robertson2009bm25}. SSR and BM25 operate
under the same maximum candidate budget and the same LLM arbitration
procedure. This controlled comparison allows the effects of maintaining
a compact candidate context to be separated from differences
attributable to the retrieval mechanism itself.

The evaluation addresses three research questions:

\begin{itemize}

    \item \textbf{RQ1:} How does full-catalog LLM orchestration scale
    with registry size, and to what extent can SSR-based candidate
    pre-selection reduce its operational overhead?

    \item \textbf{RQ2:} How effective and efficient is the SSR
    retrieval mechanism on naturally occurring text, and what are the
    contributions of IDF weighting and the ISI?

    \item \textbf{RQ3:} How does retrieval-based candidate
    pre-selection affect end-to-end agent selection, candidate context,
    cost, latency, and operational reliability as registry size grows,
    and how does SSR compare with BM25 under the same maximum candidate
    budget?

\end{itemize}

The main contributions of this work are:

\begin{itemize}

    \item We formulate scalable LLM-based agent discovery from a
    \emph{candidate-context scalability} perspective, explicitly
    separating registry-scale retrieval, retrieval computation, and
    the compact candidate context presented to the LLM orchestrator.

    \item We propose \emph{SSR}, a deterministic and training-free retrieval method that integrates field-derived structural signals, IDF-weighted matching, and selective inverted indexing into a dedicated pre-orchestration retrieval layer for Agent Card discovery. This layer produces a compact pool of top-ranked Agent Cards while preserving explicit score traceability.

    \item We introduce \emph{SSR-AgentBench}, a reproducible synthetic
    benchmark containing 1,000 procedurally generated Agent Cards and
    250 natural-language discovery queries with designated target Agent Cards for controlled registry-scaling experiments.

    \item We provide a systems-level characterization of
    candidate-context scalability through controlled registry growth,
    public-corpus retrieval, and end-to-end LLM arbitration. The
    evaluation separates general effects of candidate pre-selection
    from SSR-specific retrieval behavior while measuring retrieval
    coverage, candidate context, prompt tokens, estimated cost, latency,
    reliability, and end-to-end selection effectiveness.

\end{itemize}

\section{Related Work}

\subsection{Agent Discovery and Routing}

Large-scale agent ecosystems require mechanisms for discovering suitable
capabilities from machine-readable descriptions
\cite{wang2025ioaSurvey,chen2025ioa}. Recent work increasingly frames
this problem as retrieval or routing. One approach
\cite{guo2025agentDiscovery} combines capability announcement with
task-driven discovery and scalable indexing. Tool-to-Agent Retrieval
\cite{lumer2025tooltoagent} embeds tools and parent agents in a shared
vector space, while Agent-as-a-Graph
\cite{nizar2026agentasagraph} combines vector retrieval with explicit
graph structure. MoMA \cite{guo2025moma} and ACE-Router
\cite{yao2026acerouter} instead study learned routing over heterogeneous
models, tools, or agents.

These works establish candidate pre-selection as an important
scalability mechanism. Our focus is complementary: we formulate
\emph{candidate-context scalability} for Agent Card discovery and
propose SSR as a deterministic, training-free, field-derived retrieval
mechanism that separates registry-scale retrieval from the compact
candidate context presented to the LLM orchestrator.

\subsection{Tool and Skill Retrieval under Context Constraints}

Tool retrieval provides the closest neighboring literature. ToolRet
\cite{shi2025toolret} formulates large-scale tool retrieval as an
information-retrieval problem, PORTS \cite{molfetta2025ports} aligns
retrieval with downstream tool selection, ToolDreamer
\cite{sengupta2026tooldreamer} addresses lexical--semantic mismatch,
and ToolScope \cite{liu2026toolscope} combines tool consolidation with
context-aware filtering.

Other methods make retrieval adaptive. Dynamic Tool Dependency
Retrieval (DTDR) \cite{patel2026dtdr} updates retrieval as the
tool-calling plan evolves, ToolQP \cite{fang2026toolqp} decomposes
requests into iterative retrieval queries, and SkillRet
\cite{cho2026skillret} extends retrieval to large libraries of reusable
agent skills. Recent work on adaptive shortlist depth
\cite{repantis2026tooldepth} further treats candidate count as a
decision variable governing the trade-off between retrieval coverage
and LLM selection.

This literature shows that candidate pre-selection and retrieval-first
orchestration are established ideas. Our contribution is therefore not
the retrieve-then-shortlist pattern itself, but its explicit formulation
and systems-level characterization for Agent Card discovery.

\subsection{Retrieval Foundations}

SSR builds on established information-retrieval principles. Sparse
methods such as BM25 use lexical matching and term-weighted inverted
indexes \cite{robertson2009bm25}, whereas dense approaches such as DPR
\cite{karpukhin2020dpr} and ColBERT \cite{khattab2020colbert} learn
semantic representations that reduce dependence on exact lexical
overlap. RAG \cite{lewis2020rag} illustrates the broader retrieval-first
principle of selecting external evidence before language-model
generation.

SSR does not claim novelty for IDF weighting, sparse retrieval, or
inverted indexing. Its specific design derives deterministic unigram
and bigram signals from capability-bearing Agent Card fields, weights
matched signals with IDF, and uses an Inverted Signal Index to restrict
scoring to structurally matching candidates. The resulting mechanism is
training-free and keeps retrieval scores traceable to explicit Agent
Card evidence.

\subsection{LLM Agent Systems and Interoperability Protocols}

LLM-based agent systems study reasoning, coordination, workflow
execution, learning, and tool use
\cite{li2024llmMasSurvey,wang2024autonomousAgents,aratchige2025harmony}.
Representative systems span multi-agent coordination
\cite{wu2023autogen,li2023camel,hong2024metagpt} and reasoning,
tool-use, embodied, or learning settings
\cite{yao2023react,schick2023toolformer,wang2023voyager,
luo2025agentlightning}. The present work focuses on the preceding
systems question of which candidate capabilities should enter the LLM
context before final arbitration.

Interoperability protocols provide the machine-readable objects over
which such discovery can operate. MCP exposes structured tool
descriptions \cite{mcp2025}, while A2A represents agents through
\emph{Agent Cards} containing identity, capabilities, skills,
interfaces, and related metadata \cite{a2a2025}. These protocols define
capability representation and interaction but do not prescribe a
large-registry relevance-ranking mechanism. SSR complements them by
operating directly over field-derived Agent Card information before LLM
arbitration.

\section{Structural Signal Retriever}

The \emph{Structural Signal Retriever} (SSR) is the retrieval method
proposed in this work for Agent Card discovery. It is deterministic and
training-free, operates directly over machine-readable Agent Cards, and
produces a compact pool of top-ranked candidates for LLM arbitration.
SSR uses explicit lexical signals derived from capability-bearing Agent
Card fields, inverse document frequency (IDF) weighting, and an Inverted
Signal Index (ISI), without requiring dense representations or
supervised retrieval training.

\begin{figure*}[t]
    \centering
    \includegraphics[width=\textwidth]{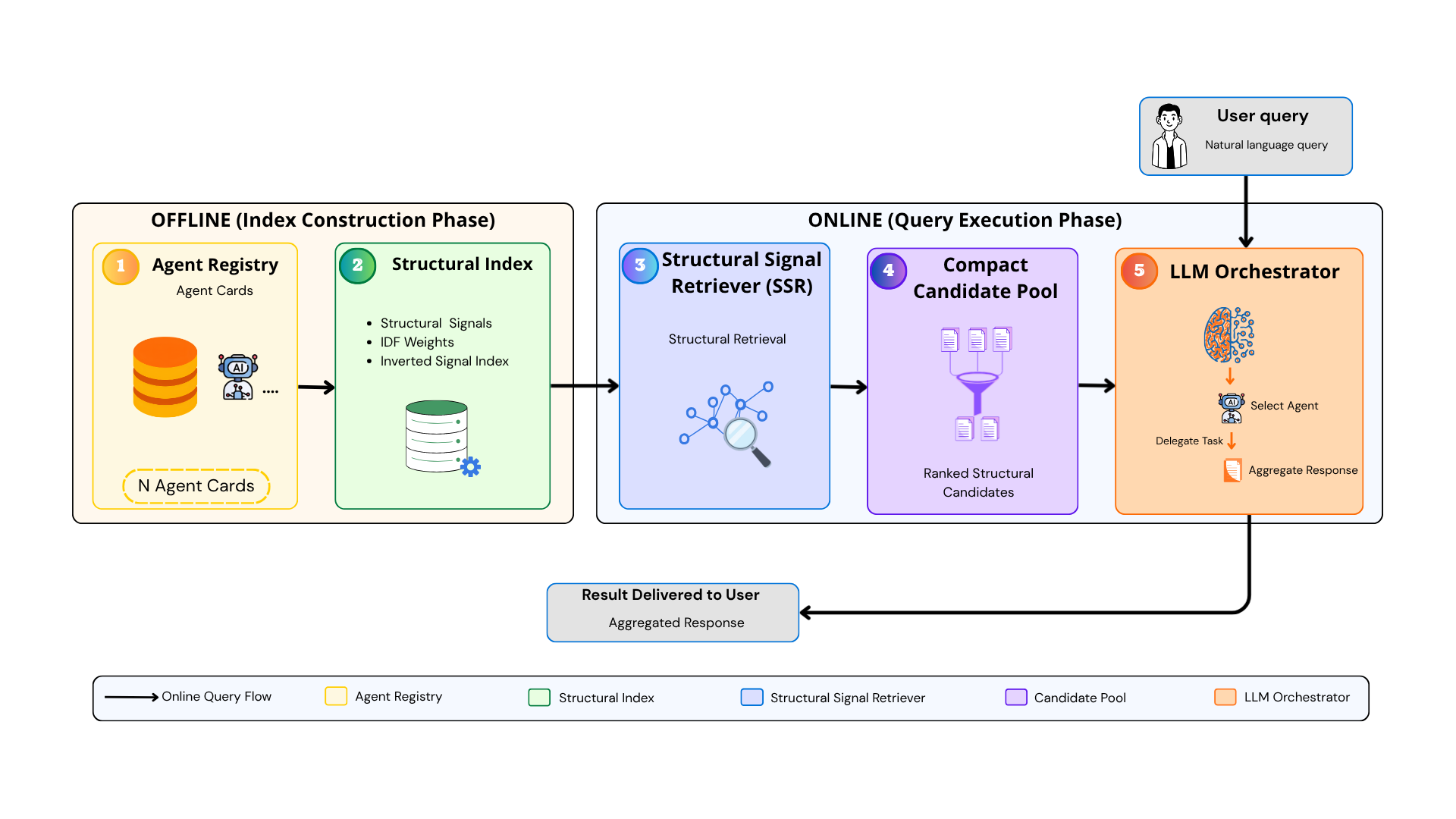}
    \caption{SSR pipeline for Agent Card discovery. Offline, the Agent
    Card registry is transformed into structural signals, IDF weights,
    and an Inverted Signal Index. Online, query signals retrieve
    structurally matching Agent Cards through the index, candidates are
    ranked by IDF-weighted signal overlap, and a compact pool of at most
    $K$ top-ranked Agent Cards is presented to the LLM orchestrator for
    final agent selection.}
    \label{fig:ssr_architecture}
\end{figure*}

In Fig.~\ref{fig:ssr_architecture}, Blocks~1--2 represent the offline registry and index-construction stages, Block~3 executes query-time SSR retrieval, Block~4 forms the compact candidate pool, and Block~5 performs final LLM arbitration. In the offline stage, the Agent Card registry is processed to extract structural signals, compute document frequencies and IDF weights, and construct the ISI. In the online stage, the user query is mapped to the same signal space; its signals address the ISI to retrieve matching Agent Cards, which are then scored and ranked. Finally, up to $K$
top-ranked Agent Cards form the compact candidate pool presented to the
LLM orchestrator for final agent selection. Thus, the registry may grow
independently of the amount of candidate information presented to the
LLM.

\subsection{Problem Formulation}

The problem addressed by SSR is to discover query-relevant Agent Cards
from a growing registry without coupling the amount of information
presented to the LLM orchestrator to the total registry size. Given a
user query, the retrieval stage must identify and rank candidate Agent
Cards while the orchestration stage receives only a compact candidate
pool. The maximum candidate budget $K$ controls the number of
top-ranked Agent Cards presented to the LLM.

Let
\[
\mathcal{A}=\{AC_1,AC_2,\ldots,AC_N\}
\]
denote a registry of $N$ Agent Cards. Let $S_a$ be the structural-signal
set of Agent Card $a$ and $S_q$ the corresponding signal set for query
$q$. SSR first identifies the Agent Cards sharing at least one
structural signal with the query,
\begin{equation}
C_q=\{a\in\mathcal{A}:S_q\cap S_a\neq\emptyset\}.
\label{eq:structural_candidates}
\end{equation}

After ranking $C_q$, the candidate pool presented to the LLM is
\begin{equation}
P_q=\operatorname{TopK}(C_q,K),
\label{eq:topk_pool}
\end{equation}
with
\begin{equation}
|P_q|=\min(K,|C_q|)\leq K.
\label{eq:pool_bound}
\end{equation}
Thus, $K$ is a maximum candidate budget rather than a required pool
size. SSR does not add non-matching Agent Cards when fewer than $K$
candidates are available; if $C_q=\emptyset$, then $P_q=\emptyset$.
The two sets therefore serve different purposes: $C_q$ contains the
Agent Cards explicitly considered for scoring, whereas $P_q$ contains
the top-ranked Agent Cards presented to the LLM orchestrator.

The SSR solution consists of two main retrieval components described
next. Section~III-B defines how Agent Cards and queries are represented
and scored through structural signals, while Section~III-C describes
how the ISI retrieves the candidates that must be scored and how the
final candidate pool is ranked.

\subsection{Signal Extraction and Scoring}

The first component of SSR defines an explicit and traceable
representation for Agent Cards and queries. A \emph{structural signal}
is a lexical unit extracted deterministically from capability-bearing
fields such as an Agent Card's name, description, declared skills,
capability tags, examples, or tool descriptions. Here,
\emph{structural} refers to signal provenance from explicit fields of
the agent representation, not to syntactic or latent semantic
structure. Because retained signals remain individually identifiable,
each retrieval score can be traced to the matched signals that
contributed to it.

Agent Cards and queries use the same deterministic normalization
procedure. Text is lowercased, alphanumeric units are extracted with
\texttt{[a-z0-9]+}, tokens shorter than two characters and fixed
stopwords are removed, and the retained sequence generates unigram and
consecutive-bigram signals. Duplicate signals are removed, so scoring
uses binary signal presence rather than term frequency.

Each representation contains at most 64 signals. If more are available,
unigrams are retained first in decreasing token-frequency order with
lexical tie-breaking; bigrams are then considered in lexicographic
order. The first 64 signals are retained. Hence
\[
|S_a|\leq64,\qquad |S_q|\leq64.
\]

Candidates are scored by IDF-weighted signal overlap,
\begin{equation}
Score(q,a)=
\sum_{s\in S_q\cap S_a}idf(s),
\label{eq:ssr_score}
\end{equation}
where
\begin{equation}
idf(s)=
\log\left(\frac{N+1}{df(s)+1}\right)+1,
\label{eq:ssr_idf}
\end{equation}
and $df(s)$ is the number of indexed Agent Cards containing $s$. Since
$df(s)\leq N$, all IDF weights are positive; consequently, rarer
signals contribute more strongly than common ones.

\subsection{Inverted Signal Index and Ranking}

The second component of SSR avoids evaluating
Eq.~\eqref{eq:ssr_score} over the complete registry whenever the query
matches only part of it. SSR constructs the Inverted Signal Index
\begin{equation}
ISI(s)=\{a\in\mathcal{A}:s\in S_a\},
\label{eq:isi}
\end{equation}
and obtains the candidate set as
\begin{equation}
C_q=\bigcup_{s\in S_q}ISI(s).
\label{eq:candidate_union}
\end{equation}

For every $a\notin C_q$, $S_q\cap S_a=\emptyset$ and therefore
$Score(q,a)=0$. Conversely, because every IDF weight is positive, every
$a\in C_q$ receives a positive score. Consequently, the ISI excludes
only zero-scoring Agent Cards and preserves the scores and relative
ordering of all positive-scoring candidates under exhaustive
Structural+IDF evaluation. When fewer than $K$ positive-scoring
candidates exist, SSR returns only those candidates rather than adding
zero-score Agent Cards. RQ2 verifies this positive-score ranking
equivalence empirically.

Candidates are ranked by descending $Score(q,a)$, with ties resolved by
descending matched-signal count and then ascending Agent Card
identifier. At most the first $K$ candidates are retained, yielding
\[
\mathcal{A}\rightarrow C_q\rightarrow P_q,
\]
where $C_q$ is the set explicitly scored after ISI lookup and $P_q$ is
the compact candidate pool presented to the LLM orchestrator.

\subsection{Retrieval Procedure and Complexity}

Algorithm~\ref{alg:ssr} presents the online SSR procedure that
implements the retrieval path shown in Fig.~\ref{fig:ssr_architecture}.
Lines~1--4 extract the query signals and use their ISI posting lists to
construct $C_q$. Lines~5--6 return an empty result when no Agent Card
shares structural evidence with the query. Lines~7--8 compute the
IDF-weighted score for each retrieved candidate, and
Lines~9--11 rank the candidates, retain up to $K$ top-ranked Agent
Cards, and return the resulting pool $P_q$.

\begin{algorithm}[t]
\caption{Structural Signal Retrieval (SSR)}
\label{alg:ssr}

\KwIn{Query $q$; Inverted Signal Index $ISI$; IDF weights; maximum
candidate budget $K$}
\KwOut{Candidate pool $P_q$}

Extract query signals $S_q$\;

$C_q\leftarrow\emptyset$\;

\ForEach{\normalfont $s\in S_q$}{
    $C_q\leftarrow C_q\cup ISI(s)$\;
}

\If{$C_q=\emptyset$}{
    \Return{$\emptyset$}\;
}

\ForEach{\normalfont $a\in C_q$}{
    $Score(q,a)\leftarrow
    \displaystyle\sum_{s\in S_q\cap S_a}idf(s)$\;
}

Rank $C_q$ by descending $Score(q,a)$, then descending
$|S_q\cap S_a|$, then ascending Agent Card identifier\;

$P_q\leftarrow\operatorname{TopK}(C_q,K)$\;

\Return{$P_q$}\;

\end{algorithm}

Let $p_s=|ISI(s)|$ denote the posting-list length of query signal $s$.
Candidate construction requires
\begin{equation}
O\left(\sum_{s\in S_q}p_s\right)
\end{equation}
expected work under constant-time set insertion. Scoring the retrieved
set requires at most
\begin{equation}
O(|C_q||S_q|)
\end{equation}
expected membership operations. Because the evaluated implementation
sorts the complete retrieved set before top-$K$ selection, ranking costs
\begin{equation}
O(|C_q|\log|C_q|).
\end{equation}
The resulting online retrieval complexity is therefore
\begin{equation}
O\left(
\sum_{s\in S_q}p_s
+
|C_q||S_q|
+
|C_q|\log|C_q|
\right).
\label{eq:ssr_complexity}
\end{equation}

Since $|S_q|\leq64$, the query representation does not grow with
registry size. Retrieval work depends on the posting lists reached by
the query and the resulting candidate set rather than requiring
Structural+IDF scoring over all $N$ Agent Cards. Independently,
Eq.~\eqref{eq:pool_bound} ensures that no more than $K$ top-ranked
Agent Cards are presented to the LLM orchestrator.

\section{Experimental Design}

The evaluation follows the three research questions introduced in
Section~I. RQ1 measures the operational effect of registry growth and
SSR-based candidate pre-selection, RQ2 evaluates SSR retrieval
effectiveness and efficiency on a large public corpus, and RQ3 compares
full-catalog orchestration with SSR and BM25 candidate pre-selection
under the same maximum candidate budget.

\subsection{SSR-AgentBench}

Public benchmarks exist for large-scale tool retrieval and MCP-oriented tool use \cite{mo2025livemcpbench,dong2026etom,wang2026mcpflow}, but, to the best of our knowledge, no public benchmark currently combines
structured A2A-style Agent Cards, natural-language discovery queries, reproducible query-to-agent references, and controlled registry scaling. We therefore introduce \emph{SSR-AgentBench}, a synthetic benchmark for controlled agent-discovery experiments.

SSR-AgentBench contains 1,000 Agent Cards, including 850 base Agent Cards and 150 capability twins, together with 250 English discovery queries. Capability twins preserve the core domain and specialty of a base profile while varying query-relevant operational attributes. Each query is paired with a designated target whose complete query-relevant attribute combination is unique under the controlled generator,
supporting deterministic evaluation without assuming that real
registries contain only one acceptable agent.

\begin{figure*}[t]
    \centering
    \includegraphics[scale=0.40]{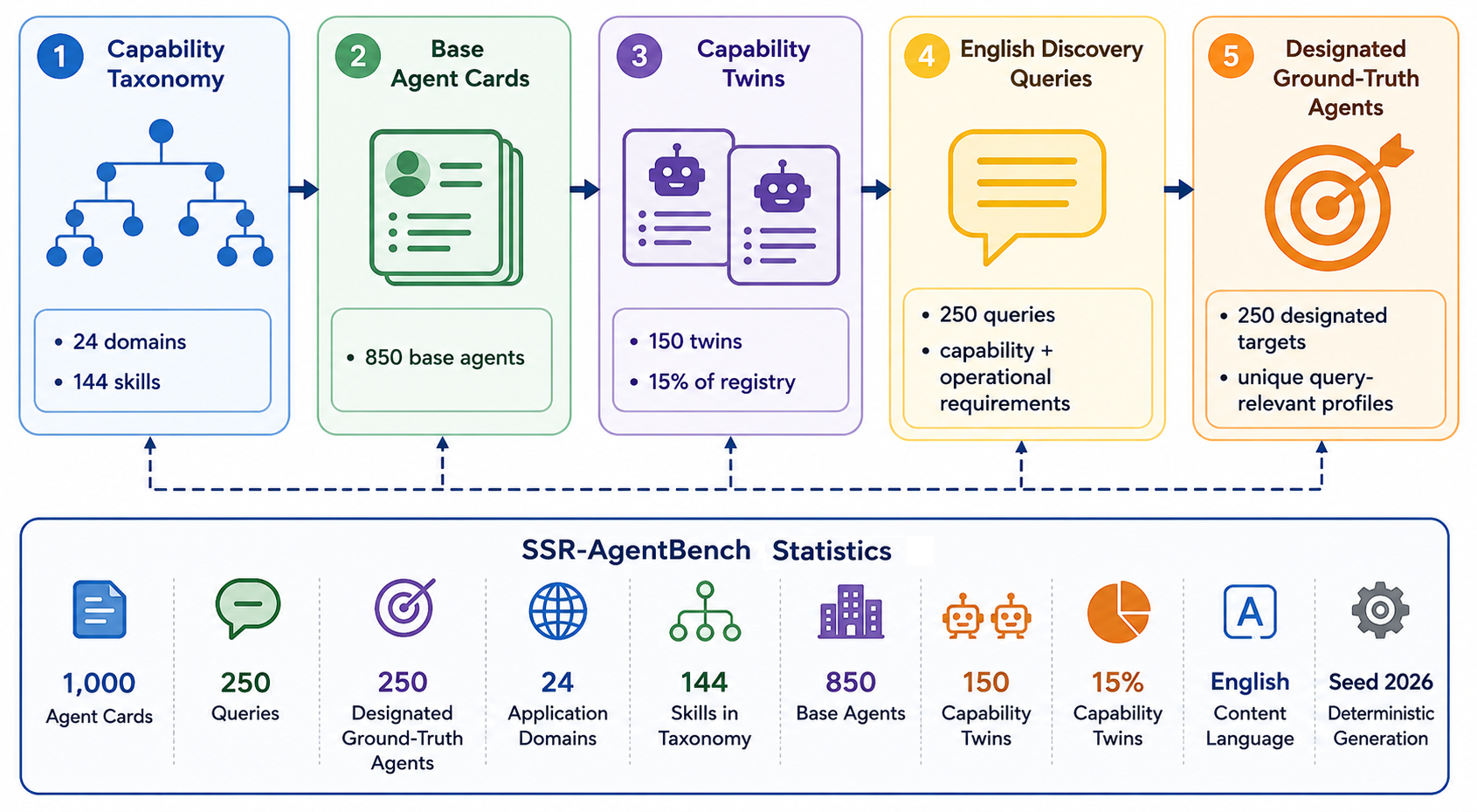}
    \caption{SSR-AgentBench generation pipeline. A taxonomy of 24
    application domains and 144 skills produces 850 base Agent Cards;
    150 capability twins introduce controlled capability overlap.
    Natural-language queries are generated from capability and
    operational profiles and paired with designated target Agent Cards.}
    \label{fig:AgentBench}
\end{figure*}

Figure~\ref{fig:AgentBench} summarizes the benchmark-generation
pipeline. Block~1 defines the capability taxonomy of 24 domains and 144 skills; Block~2 generates the 850 base Agent Cards; Block~3 adds 150 capability twins to introduce controlled capability overlap;
Block~4 produces 250 English discovery queries from capability and operational requirements; and Block~5 assigns the corresponding designated target Agent Cards. The lower panel summarizes the resulting benchmark statistics, also reported in
Table~\ref{tab:benchmark_statistics}. Generation uses seed 2026, and registry scaling relies on deterministic nested subsets rather than regenerating Agent Cards at each size.

\begin{table}[t]
\centering
\caption{SSR-AgentBench statistics.}
\label{tab:benchmark_statistics}
\begin{tabular}{lr}
\toprule
\textbf{Component} & \textbf{Value} \\
\midrule
Agent Cards & 1000 \\
Base / capability twins & 850 / 150 \\
Queries & 250 \\
Application domains & 24 \\
Capability taxonomy & 144 skills \\
Language & English \\
Generation seed & 2026 \\
\bottomrule
\end{tabular}
\end{table}

\subsection{RQ1: Registry-Scaling Overhead}

RQ1 evaluates how full-catalog LLM orchestration scales with registry
size and whether SSR-based candidate pre-selection can maintain a
compact candidate context. All 250 SSR-AgentBench queries are evaluated
at
\[
N\in\{10,100,500,1000\}.
\]
For each query, deterministic nested registries are constructed such
that
\[
\mathcal{A}_{10}\subset
\mathcal{A}_{100}\subset
\mathcal{A}_{500}\subset
\mathcal{A}_{1000},
\]
with the designated target present at every size.

Two configurations are compared: \emph{full catalog}, which forwards
every Agent Card to the LLM, and \emph{SSR pre-selection}, which
retrieves at most $K=10$ candidates before arbitration. Both use the
same LLM candidate projection: a one-based index, Agent Card name, and
description. SSR does not add non-matching Agent Cards when fewer than
$K$ candidates are available.

RQ1 reports candidate count, provider-reported input tokens, estimated
inference cost, and online end-to-end latency; candidate recall and final
selection accuracy are retained as diagnostics. SSR latency includes
query-time retrieval, prompt construction, the LLM request, and response
parsing, while registry construction and indexing are excluded. The two
configurations are executed in deterministically alternating order
across query--registry pairs.

\subsection{RQ2: Retrieval Effectiveness and Efficiency}

RQ2 evaluates SSR on naturally occurring text using the MS MARCO
Passage Ranking collection \cite{bajaj2016msmarco}, which contains
8,841,823 passages. Because MS MARCO does not provide native Agent
Cards, each passage is treated as an abstract document-derived Agent
Card, each query as an agent-discovery request, and positively judged
passages as relevant Agent Cards under this abstraction. All 6,980
development queries with at least one positive relevance judgment are
used for effectiveness evaluation.

Retrieval depth is fixed at $K=10$, and effectiveness is measured using
Recall@1, Recall@3, Recall@5, Recall@10, and MRR@10. Five configurations
are evaluated:

\begin{enumerate}
    \item \textbf{TF-IDF}, using an unnormalized lexical dot product;
    \item \textbf{BM25} \cite{robertson2009bm25}, over the same normalized
    unigram representation;
    \item \textbf{Structural Overlap + ISI}, using equal weights for
    matched structural signals;
    \item \textbf{Structural + IDF (full scan)}, using SSR scoring without
    the ISI; and
    \item \textbf{SSR}, combining field-derived signals, IDF weighting,
    and the ISI.
\end{enumerate}

Structural Overlap + ISI isolates the contribution of IDF weighting,
whereas Structural + IDF (full scan) isolates the contribution of the
ISI. Effectiveness for indexed methods is evaluated on all 6,980
eligible queries. Online latency and direct exhaustive Structural+IDF
execution are measured on a common deterministic 200-query subset
sampled with seed 42.

For each query in this subset, exhaustive Structural+IDF scores all
8,841,823 passages, discards zero-score results, and returns the
unpadded top-$K$ positive-scoring candidates. As established in
Section~III, the ISI excludes only zero-scoring candidates; direct
comparison with SSR therefore verifies top-$K$ agreement while
quantifying the computational cost of exhaustive scoring. Index
construction and IDF computation are treated as offline operations.

\subsection{RQ3: End-to-End Agent Discovery}

RQ3 evaluates the complete LLM selection pipeline at
\[
N\in\{100,500,1000,5000\}
\]
using 200 MS MARCO development queries sampled uniformly with seed 42.
Sampling occurs before SSR retrieval and is independent of SSR ranks,
candidate pools, and LLM outcomes.

For each query, relevant document-derived Agent Cards are retained and
deterministic distractor Agent Cards form nested registries,
\[
\mathcal{A}_{100}\subset
\mathcal{A}_{500}\subset
\mathcal{A}_{1000}\subset
\mathcal{A}_{5000}.
\]
Three configurations are evaluated: full-catalog orchestration, SSR
candidate pre-selection, and BM25 candidate pre-selection. SSR and BM25
use the same maximum candidate budget, query-specific registries, LLM
prompt, and parser. Neither method pads non-matching results or falls
back to the full catalog. If the retrieved pool is empty, no LLM call
is made and the case remains an unsuccessful end-to-end outcome.

\subsubsection{Candidate-Budget Selection and Independent Validation}

The original RQ3 protocol evaluated SSR retrieval sensitivity for
\[
K\in\{5,10,20,50\}
\]
on the same 200-query sample later used for end-to-end evaluation,
without using LLM selection outcomes. The reported end-to-end
experiments use $K=20$, with the same maximum budget applied to BM25.

To assess whether this operating point was sample-specific, the 200 RQ3
queries were excluded and 1,000 queries were sampled uniformly from the
remaining 6,780 using seed 2027. The validation reused the same nested
registry sizes, SSR configuration, and candidate budgets, and made no
LLM calls. The decision rule selected the smallest $K$ whose candidate
recall was within 0.02 absolute of the best tested recall at every
registry size:
\[
K^*=
\min\left\{
K:
R(N,K)\geq
\max_{K'}R(N,K')-0.02,\ \forall N
\right\}.
\]
The independent validation selected $K=20$, corroborating the operating
point used in the reported end-to-end experiments. Because this
validation was retrospective, a fully prospective protocol would select
and freeze $K$ before evaluation on a disjoint end-to-end test set.

\subsubsection{Effectiveness and Paired Statistical Analysis}

For candidate pre-selection, \emph{candidate recall} is the fraction of
query--registry pairs for which at least one relevant candidate is
retrieved. \emph{Conditional selection accuracy} is the fraction of
retrieval-hit cases in which the LLM selects a relevant candidate, and
\emph{end-to-end accuracy} is the fraction of all query--registry pairs
for which the complete pipeline returns a relevant candidate. Retrieval
misses, empty pools, API failures, invalid indices, and irrelevant
selections remain end-to-end failures.

For each $N$, 95\% Wilson score intervals are reported for marginal
end-to-end accuracy. Because SSR and BM25 are evaluated on the same 200
queries and query-specific registries, their primary paired comparison
uses a two-sided exact McNemar test on query-level
\texttt{end\_to\_end\_correct} outcomes, with Holm correction across the
four registry sizes. Candidate recall is analyzed analogously as a
secondary paired outcome. Nonsignificant results are interpreted as no
statistically detectable difference under the evaluated sample, not as
evidence of statistical equivalence.

\subsection{Implementation and Reproducibility}

\textbf{LLM configuration.}
RQ1 and RQ3 use \texttt{gpt-4o-mini}, temperature 0, and JSON-mode
responses. Candidates are represented by sequential one-based indices,
and the returned \texttt{selected\_index} is validated and mapped to
the underlying Agent Card or passage identifier. Accounting uses fixed
rates of \$0.15 per million input tokens and \$0.60 per million output
tokens.

\textbf{Retrieval implementation.}
SSR retains at most 64 structural signals per representation and uses
the normalization and IDF definition in Section~III. BM25 is implemented
with SQLite FTS5 \texttt{bm25()} over the same normalized unigram
tokenization used for indexed documents and queries; lower FTS5 scores
rank first, with passage identifier used for deterministic tie-breaking.
The implementation uses $k_1=1.2$ and $b=0.75$. Offline index
construction is excluded from reported online latency for both methods.

\textbf{Execution environment.}
The full-scale RQ2 experiment was executed on an Amazon SageMaker
\texttt{ml.m5d.8xlarge} instance running Ubuntu 22.04.5 LTS with 32
logical CPUs, approximately 124~GiB RAM, an Intel Xeon Platinum 8259CL
2.50~GHz processor, Python 3.12.9, and SQLite 3.48.0.

\textbf{Token, cost, and latency accounting.}
Successful LLM calls use provider-reported input/output token counts.
Estimated cost is
\[
\widehat{C}=
\frac{T_{\mathrm{in}}}{10^6}p_{\mathrm{in}}+
\frac{T_{\mathrm{out}}}{10^6}p_{\mathrm{out}}.
\]
In RQ3, when provider usage is unavailable, input tokens are estimated
from prompt length as
\[
\widehat{T}_{\mathrm{in}}=
\max\left(1,
\left\lfloor\frac{|\mathrm{prompt}|}{4}\right\rfloor
\right),
\]
and the resulting cost is treated as a prompt-based estimate rather
than measured inference usage.

RQ1 SSR latency includes online retrieval, prompt construction, the LLM
request, and response processing; full-catalog latency excludes
retrieval. RQ2 latency measures query-time retrieval only. In RQ3,
candidate-pre-selection latency includes retrieval, prompt construction,
the LLM request, and response processing, while full-catalog latency
includes prompt construction and the LLM request. Failed requests
contribute to reliability and end-to-end effectiveness but not to
successful orchestration latency. Because the BM25 RQ3 run was executed
separately, SSR--BM25 latency differences are interpreted descriptively
rather than as paired comparisons.

\textbf{Determinism and artifacts.}
SSR-AgentBench and RQ1 use seed 2026; RQ2 timing/equivalence and the
RQ3 end-to-end sample use seed 42; independent $K$ validation uses seed
2027. The RQ3 retrievers operate on the same reconstructed
query-specific registries. Experiment scripts record configuration
metadata and per-query outputs for auditing and reproduction. The
benchmark generator and evaluation code are available at
\url{https://github.com/gmaldonadoc/sscr-agent-routing}.

\section{Results}

Results are reported by research question, separating registry-scaling
overhead, retrieval behavior, and end-to-end LLM arbitration.

\subsection{RQ1: Registry-Scaling Overhead}

\begin{table*}[t]
\centering
\caption{Scaling evaluation on SSR-AgentBench. SSR uses a maximum
candidate budget of $K=10$. ``Agent Cards'' is the mean number of
candidate representations included in the LLM prompt.}
\label{tab:scaling}
\scriptsize
\setlength{\tabcolsep}{7pt}
\begin{tabular*}{\textwidth}{@{\extracolsep{\fill}}llrrrr@{}}
\toprule
Agents & Method & Agent Cards & Prompt Tokens & Est. Cost (USD) & Latency (s) \\
\midrule
10   & Full catalog & 10.0   & 771    & 0.000120 & 0.722 \\
10   & SSR          & 10.0   & 771    & 0.000120 & 0.715 \\
\midrule
100  & Full catalog & 100.0  & 6,481  & 0.000976 & 1.165 \\
100  & SSR          & 10.0   & 769    & 0.000120 & 0.748 \\
\midrule
500  & Full catalog & 500.0  & 31,890 & 0.004788 & 2.423 \\
500  & SSR          & 10.0   & 767    & 0.000119 & 0.779 \\
\midrule
1000 & Full catalog & 1000.0 & 63,645 & 0.009551 & 2.647 \\
1000 & SSR          & 10.0   & 767    & 0.000119 & 0.769 \\
\bottomrule
\end{tabular*}
\end{table*}

Across 250 queries, four registry sizes, and two orchestration
configurations, all 2,000 LLM calls complete and parse successfully.
Table~\ref{tab:scaling} shows that full-catalog prompt size grows with
the registry, reaching 63,645 mean input tokens at $N=1000$, whereas SSR
uses 767 tokens with at most 10 Agent Cards. Relative to full catalog,
SSR reduces mean input tokens by 88.1\%, 97.6\%, and 98.8\% at
$N=100$, 500, and 1000, respectively. At $N=1000$, estimated cost also
falls by 98.8\%, from \$0.009551 to \$0.000119 per query, and mean
successful end-to-end latency decreases from 2.647~s to 0.769~s
(70.9\%, or $3.44\times$).

As diagnostics, SSR candidate recall and final selection accuracy are
1.000 at all four registry sizes. Full-catalog selection accuracy is
1.000 at $N=10$ and 100 and 0.976 at $N=500$ and 1000; these
differences are not interpreted as comparative effectiveness evidence.

\textbf{Answer to RQ1.} Full-catalog orchestration scales prompt size,
cost, and latency with registry size, whereas SSR-based candidate
pre-selection maintains a compact LLM candidate context and
substantially reduces operational overhead.

\subsection{RQ2: Retrieval Effectiveness and Efficiency}

\begin{table*}[t]
\centering
\caption{Retrieval effectiveness and online efficiency on MS MARCO.
Effectiveness uses all 6,980 eligible queries; latency uses a common
200-query subset. ``Avg. Passages Scored`` counts passages explicitly
evaluated by the structural scoring function.}
\label{tab:msmarco_retrieval}
\scriptsize
\setlength{\tabcolsep}{4.5pt}
\begin{tabular}{lccccccc}
\toprule
Method & R@1 & R@3 & R@5 & R@10 & MRR@10 & Mean Lat. (ms) & Avg. Passages Scored \\
\midrule
TF-IDF (dot product)
& 0.0087 & 0.0266 & 0.0398 & 0.0619 & 0.0221 & 906.9 & -- \\
BM25
& \textbf{0.1026} & \textbf{0.2255} & \textbf{0.2923}
& \textbf{0.3890} & \textbf{0.1829} & 822.6 & -- \\
Structural Overlap + ISI
& 0.0798 & 0.1768 & 0.2351 & 0.3209 & 0.1462 & 937.2 & 286,638 \\
Structural + IDF (full scan)
& 0.0834 & 0.1872 & 0.2510 & 0.3378 & 0.1544 & 69,848.3 & 8,841,823 \\
\textbf{SSR (Structural + IDF + ISI)}
& 0.0834 & 0.1872 & 0.2510 & 0.3378 & 0.1544 & 941.5 & 286,638 \\
\bottomrule
\end{tabular}
\end{table*}

Table~\ref{tab:msmarco_retrieval} shows that BM25 is the strongest raw
retrieval baseline, reaching R@10=0.3890 and MRR@10=0.1829 versus
0.3378 and 0.1544 for SSR. SSR therefore does not establish general
retrieval superiority over established lexical methods.

Within the SSR representation, IDF weighting consistently improves all
reported ranking metrics relative to Structural Overlap + ISI, including
R@10 by 5.3\% and MRR@10 by 5.6\%. The ISI affects computation rather
than the positive-scoring ranking: direct exhaustive execution on the
common 200-query subset verifies exact unpadded top-$K$ agreement with
full-scan Structural+IDF. Exhaustive scoring evaluates all 8,841,823
passages and requires 69,848.3~ms per query on average, whereas SSR
explicitly scores approximately 286,638 passages in 941.5~ms. This
corresponds to 96.8\% fewer explicitly scored passages and a
$74.2\times$ mean retrieval-latency speedup.

\textbf{Answer to RQ2.} IDF improves ranking effectiveness within the
structural representation, while the ISI preserves the positive-scoring
ranking and provides the principal computational reduction.

\subsection{RQ3: End-to-End Agent Discovery}

RQ3 compares full-catalog orchestration with SSR and BM25 candidate
pre-selection under the same maximum candidate budget $K=20$ on 200
independently sampled MS MARCO queries and registries of
\[
N\in\{100,500,1000,5000\}.
\]

\subsubsection{Candidate-Budget Robustness}

On the 200 RQ3 queries, SSR candidate recall with $K=20$ is 0.980 for
$N\leq1000$ and 0.950 at $N=5000$. Increasing the budget to $K=50$
raises recall at $N=5000$ only to 0.960 while substantially increasing
the candidate pool.

Independent retrieval-only validation on 1,000 disjoint queries
corroborates the chosen operating point. At $K=20$, candidate recall is
0.973, 0.969, 0.965, and 0.941 for $N=100$, 500, 1000, and 5000,
respectively, satisfying the predefined within-0.02 decision rule at
every registry size. Only $K=20$ and $K=50$ satisfy the rule, so the
smallest qualifying budget is $K=20$. At $N=5000$, increasing to
$K=50$ improves recall by 0.018 while increasing mean pool size from
17.822 to 41.311 ($2.318\times$). The validation supports the robustness
of $K=20$ but remains retrospective rather than a prospective held-out
selection protocol.

\subsubsection{Retrieval and Selection Decomposition}

\begin{table}[t]
\centering
\caption{Candidate-pre-selection effectiveness decomposition. Both
methods use maximum candidate budget $K=20$.}
\label{tab:rq3_decomposition}
\footnotesize
\setlength{\tabcolsep}{4pt}
\begin{tabular}{lrccc}
\toprule
\textbf{Method} & $N$ & \textbf{Cand. Recall} & \textbf{Cond. Acc.} & \textbf{E2E Acc.} \\
\midrule
SSR  & 100  & 0.980 & 1.0000 & 0.980 \\
BM25 & 100  & 0.980 & 1.0000 & 0.980 \\
\midrule
SSR  & 500  & 0.980 & 0.9898 & 0.970 \\
BM25 & 500  & 0.980 & 0.9796 & 0.960 \\
\midrule
SSR  & 1000 & 0.980 & 0.9796 & 0.960 \\
BM25 & 1000 & 0.980 & 0.9898 & 0.970 \\
\midrule
SSR  & 5000 & 0.950 & 0.9316 & 0.885 \\
BM25 & 5000 & 0.970 & 0.9227 & 0.895 \\
\bottomrule
\end{tabular}
\end{table}

Table~\ref{tab:rq3_decomposition} shows comparable observed
end-to-end effectiveness for SSR and BM25 candidate pre-selection. For
$N\leq1000$, candidate recall is 0.980 for both methods. At $N=5000$,
BM25 has higher candidate recall (0.970 vs. 0.950), while SSR has
slightly higher conditional selection accuracy (0.9316 vs. 0.9227),
yielding E2E accuracies of 0.895 and 0.885, respectively. Retrieval
coverage and LLM arbitration therefore contribute separately to
end-to-end error.

Paired exact McNemar tests on \texttt{end\_to\_end\_correct} detect no
statistically significant difference between SSR and BM25 at any
registry size after Holm correction (adjusted $p=1.000$ throughout).
Candidate recall likewise shows no statistically detectable paired
difference after correction; at $N=5000$, the adjusted $p$-value is
0.500. These results support describing observed effectiveness as
comparable under the evaluated sample, but do not establish statistical
equivalence. The 95\% Wilson intervals for SSR E2E accuracy are
$[0.950,0.992]$, $[0.936,0.986]$, $[0.923,0.980]$, and
$[0.833,0.922]$ as $N$ increases; for BM25 they are
$[0.950,0.992]$, $[0.923,0.980]$, $[0.936,0.986]$, and
$[0.845,0.930]$.

\subsubsection{Candidate Context and Operational Behavior}

\begin{table*}[t]
\centering
\caption{End-to-end infrastructure benchmark on 200 RQ3 queries.
SSR and BM25 use $K=20$. API success is computed over attempted calls;
latency is reported only for successful executions. Asterisks denote
prompt-based estimates for failed full-catalog requests at $N=5000$.}
\label{tab:end_to_end}
\scriptsize
\setlength{\tabcolsep}{3.8pt}
\begin{tabular*}{\textwidth}{@{\extracolsep{\fill}}llrrrrrrr@{}}
\toprule
Method & $N$ & E2E Acc. & Attempt & API Success & Pool & Tokens &
Est. Cost (USD) & Latency (s) \\
\midrule
SSR & 100 & 0.980 & 1.000 & 1.000 & 4.34 & 444 & 0.000072 & 0.674 \\
BM25 & 100 & 0.980 & 0.985 & 1.000 & 4.34 & 444 & 0.000072 & 0.620 \\
Full catalog & 100 & 1.000 & 1.000 & 1.000 & 100 & 8,154 & 0.001227 & 1.099 \\
\midrule
SSR & 500 & 0.970 & 1.000 & 1.000 & 12.06 & 1,125 & 0.000175 & 0.643 \\
BM25 & 500 & 0.960 & 0.990 & 1.000 & 12.06 & 1,078 & 0.000167 & 0.628 \\
Full catalog & 500 & 0.990 & 1.000 & 1.000 & 500 & 40,705 & 0.006110 & 2.449 \\
\midrule
SSR & 1000 & 0.960 & 1.000 & 1.000 & 15.14 & 1,386 & 0.000214 & 0.666 \\
BM25 & 1000 & 0.970 & 0.990 & 1.000 & 15.14 & 1,318 & 0.000204 & 0.646 \\
Full catalog & 1000 & 0.955 & 1.000 & 1.000 & 1000 & 81,368 & 0.012209 & 3.915 \\
\midrule
SSR & 5000 & 0.885 & 1.000 & 1.000 & 18.67 & 1,690 & 0.000260 & 0.686 \\
BM25 & 5000 & 0.895 & 0.990 & 1.000 & 18.67 & 1,561 & 0.000241 & 0.650 \\
Full catalog & 5000 & 0.000 & 1.000 & 0.000 & 5000 &
447,761$^{*}$ & 0.067164$^{*}$ & -- \\
\bottomrule
\end{tabular*}
\end{table*}

Table~\ref{tab:end_to_end} shows that full-catalog context grows rapidly with registry size: mean input tokens increase from 8,154 at $N=100$ to 81,368 at $N=1000$, while mean successful latency rises from 1.099~s to 3.915~s. At $N=5000$, the estimated prompt size is 447,761 tokens and none of the 200 attempted LLM calls completes successfully; the reported token and cost values at this size are prompt-based estimates.

In contrast, SSR and BM25 candidate pre-selection maintain compact candidate contexts as the registry grows. At $N=5000$, both present 18.67 candidates on average; SSR uses 1,690 mean input tokens and BM25 1,561. Relative to full catalog, SSR reduces mean prompt size by 94.6\%, 97.2\%, 98.3\%, and 99.6\% for $N=100$, 500, 1000, and 5000,
respectively. BM25 shows the same context-scaling behavior, indicating that the dominant reduction arises from candidate pre-selection rather than SSR scoring alone.

BM25 produces empty pools for a small number of query--registry pairs, so its LLM-attempt rate is 0.985 at $N=100$ and 0.990 thereafter. All attempted SSR and BM25 LLM calls complete successfully. Their latency values are interpreted descriptively because the BM25 run was executed separately.

\textbf{Answer to RQ3.} Candidate pre-selection keeps the LLM candidate context compact and avoids the operational failure observed for
full-catalog orchestration at $N=5000$. SSR and BM25 show comparable observed end-to-end effectiveness under the same maximum candidate budget, while SSR provides the deterministic, training-free, field-derived, and score-traceable retrieval mechanism proposed in this work.

\section{Discussion}

The results support treating scalable agent discovery as a joint
retrieval--orchestration problem. Full-catalog orchestration couples
registry size directly to the amount of information presented to the
LLM, whereas candidate pre-selection separates registry-scale retrieval
from the compact candidate context used for final arbitration. RQ3
shows that this systems effect is not specific to SSR: both SSR and BM25
maintain small candidate pools under the same maximum candidate budget,
while full-catalog context grows rapidly with registry size.

This distinction is central to the interpretation of SSR. RQ2 shows
that BM25 achieves higher raw retrieval effectiveness on the MS MARCO
adaptation, so SSR should not be viewed as a general replacement for
established lexical retrieval. Instead, SSR provides a deterministic,
training-free, field-derived retrieval mechanism for Agent Cards.
Within this mechanism, IDF weighting improves ranking effectiveness,
the Inverted Signal Index (ISI) reduces explicit structural scoring,
and the maximum candidate budget $K$ controls the amount of candidate
information presented to the LLM orchestrator. These objectives are
related but distinct.

\subsection{Candidate-Context Scalability and the Role of $K$}

The principal systems property studied here is
\[
|P_q|\leq K,
\]
which places an explicit upper bound on the number of Agent Cards
presented to the LLM independently of the total registry size. This
does not imply registry-independent retrieval cost: SSR and BM25 still
query indexes whose contents and posting distributions depend on
registry composition. The bound applies to the final candidate pool,
not to all upstream retrieval work.

RQ3 also exposes the trade-off between candidate context and retrieval
coverage. At $N=5000$, SSR and BM25 achieve candidate recall of 0.950
and 0.970 and end-to-end accuracy of 0.885 and 0.895, respectively.
Paired McNemar tests detect no statistically significant difference in
end-to-end correctness after Holm correction, supporting a description
of comparable observed effectiveness without establishing statistical
equivalence.

The fixed $K=20$ operating point should be interpreted as an evaluated
systems choice rather than a universal optimum. Independent
retrieval-only validation on 1,000 disjoint queries also selected
$K=20$ under the predefined within-0.02 recall rule, supporting its
robustness. Because this validation was retrospective, however, a fully
prospective protocol should select and freeze $K$ before evaluation on
a disjoint end-to-end test set.

\subsection{Positioning Relative to Retrieval and Routing}

Candidate pre-selection and retrieval-first orchestration are
established ideas in tool and agent systems. ToolRet
\cite{shi2025toolret} studies large-scale tool retrieval; ToolScope
\cite{liu2026toolscope} reduces tool context through consolidation and
filtering; Tool-to-Agent Retrieval \cite{lumer2025tooltoagent} and
Agent-as-a-Graph \cite{nizar2026agentasagraph} exploit tool--agent
relationships; and ACE-Router \cite{yao2026acerouter} and AgentRouter
\cite{zhang2026agentrouter} study learned routing. Adaptive shortlist
depth further treats candidate count itself as a decision variable
\cite{repantis2026tooldepth}.

The contribution of this work is therefore not candidate pre-selection,
IDF weighting, or inverted indexing in isolation. Rather, we formulate
\emph{candidate-context scalability} for Agent Card discovery, propose
SSR as an Agent-Card-native retrieval mechanism, and characterize how
retrieval coverage, retrieval computation, candidate context, cost,
latency, reliability, and LLM arbitration evolve under controlled
registry growth. BM25 candidate pre-selection provides an important
control because it separates the general effect of maintaining a compact
candidate context from properties specific to SSR.

Table~\ref{tab:comparison} summarizes this conceptual positioning
relative to representative retrieval and routing approaches.

\begin{table*}[t]
\centering
\caption{Conceptual positioning of representative approaches. The table
summarizes differences in formulation, retrieval mechanism, and
candidate control; it does not claim architectural precedence or
cross-paper performance superiority.}
\label{tab:comparison}
\scriptsize
\setlength{\tabcolsep}{4pt}

\begin{tabularx}{\textwidth}{
>{\raggedright\arraybackslash}p{2.5cm}
>{\raggedright\arraybackslash}p{2.7cm}
>{\raggedright\arraybackslash}X
>{\raggedright\arraybackslash}X
>{\raggedright\arraybackslash}X}
\toprule
\textbf{Approach} &
\textbf{Primary setting} &
\textbf{Representation / mechanism} &
\textbf{Candidate control} &
\textbf{Relation to this work} \\
\midrule

Tool-to-Agent Retrieval
\cite{lumer2025tooltoagent} &
Tool and agent retrieval &
Shared vector representations with explicit tool--agent relations &
Retrieval-ranked subset &
Dense relational alternative for tool-to-agent discovery \\

Agent-as-a-Graph
\cite{nizar2026agentasagraph} &
Tool and agent retrieval &
Vector retrieval plus explicit graph relationships &
Graph- and ranking-derived subset &
Uses richer relational structure than SSR \\

ToolScope
\cite{liu2026toolscope} &
Large-scale LLM tool use &
Tool merging and context-aware filtering &
Context-dependent filtered toolset &
Shares the goal of reducing LLM candidate context \\

ACE-Router
\cite{yao2026acerouter} &
MCP and agent routing &
Learned history-aware routing &
Router-selected candidates &
Learned, context-dependent alternative \\

AgentRouter
\cite{zhang2026agentrouter} &
Collaborative multi-agent QA &
Knowledge-graph-guided supervised routing &
Learned routing distribution &
Models contextual and relational agent structure \\

Adaptive shortlist depth \cite{repantis2026tooldepth} &
LLM tool selection &
Query-dependent shortlist-depth policy &
Adaptive depth &
Directly studies candidate-depth selection \\

SSR (this work) &
Agent discovery before LLM arbitration &
Deterministic field-derived lexical signals, IDF, and ISI &
Explicit maximum budget $K$ &
Training-free instantiation of candidate-context scalability \\

\bottomrule
\end{tabularx}
\end{table*}

SSR also provides score-level traceability: matched structural signals
and their IDF contributions remain identifiable for each retrieved Agent
Card. This property applies to the retrieval stage and should not be
interpreted as making subsequent LLM arbitration explainable.

\subsection{Limitations}

Several limitations qualify the conclusions. First, RQ2 and RQ3 use a
document-derived Agent Card abstraction. MS MARCO provides natural text
and large-scale relevance judgments, but not native Agent Cards,
structured skills, interfaces, authentication requirements, or
operational metadata. The results therefore characterize retrieval and
candidate-context behavior under this abstraction rather than in a
production agent registry.

Second, SSR-AgentBench is synthetic and uses one designated target per
query. Its controlled construction supports reproducible registry
scaling but does not capture the full heterogeneity, documentation
quality, capability evolution, access constraints, or multiple valid
agents found in operational ecosystems.

Third, SSR deliberately relies on explicit lexical unigrams and
bigrams. Vocabulary mismatch can reduce recall, and the MS MARCO
results show that this representation does not outperform BM25. More
expressive field-derived, semantic, or hybrid retrieval may improve
coverage.

Fourth, although independent validation supports $K=20$, the original
sensitivity analysis and end-to-end evaluation reuse the same 200 RQ3
queries. A cleaner prospective design would select and freeze $K$ on
validation data and evaluate it once on a disjoint end-to-end test set.

Finally, RQ3 uses 200 queries and one LLM/API configuration, and the
BM25 run was executed separately from the SSR/full-catalog run.
Absolute latency, token usage, cost, and the observed full-catalog
failure point should therefore not be generalized across models,
providers, prompts, or context limits. Reported online latency also
excludes offline index construction, so preprocessing, storage, and
maintenance costs remain outside the present evaluation.

\subsection{Future Directions}

Future work should first investigate query-adaptive candidate depth,
using score concentration, margins, or other confidence signals to
choose $K(q)$ while preserving a hard maximum candidate budget
\cite{repantis2026tooldepth}. Second, SSR could provide inexpensive
candidate generation followed by semantic retrieval or reranking when
lexical evidence is weak; ToolDreamer illustrates one semantic
reformulation strategy \cite{sengupta2026tooldreamer}. Third, retrieval
could become execution-aware, recomputing queries or candidate pools as
a plan evolves, as suggested by DTDR and ToolQP
\cite{patel2026dtdr,fang2026toolqp}.

Evaluation should also move toward native public Agent Card and skill
collections, multi-relevance judgments, prospective validation/test
splits, and dynamic registry updates. Large distributed deployments
further motivate incremental, authorization-aware, and federated
indexing.

Overall, the results support candidate-context scalability as a useful
systems perspective: retrieval coverage, retrieval computation, and the
compact candidate context presented to the LLM can be treated as
separate control points. BM25 shows that the main context-scaling
benefit is a general effect of candidate pre-selection, while SSR
contributes a deterministic, training-free, field-derived, and
score-traceable retrieval mechanism for Agent Card discovery.

\section{Conclusion}

This work studied \emph{candidate-context scalability} in LLM-based
agent discovery: the separation between registry-scale retrieval and the
compact candidate context presented to the LLM orchestrator. We propose
the \emph{Structural Signal Retriever} (SSR), a deterministic,
training-free method for Agent Card discovery based on field-derived
lexical signals, inverse document frequency (IDF), and an Inverted
Signal Index (ISI).

The three experiments provide complementary evidence. On
SSR-AgentBench, SSR reduces mean LLM input by 98.8\% and mean successful
orchestration latency by 70.9\% at $N=1000$ relative to full-catalog
orchestration. On the MS MARCO adaptation, IDF improves ranking
effectiveness within the structural representation, while the ISI
preserves the positive-scoring Structural+IDF ranking and reduces
explicitly scored passages by 96.8\%, yielding a $74.2\times$ mean
retrieval-latency speedup over exhaustive scoring. BM25 achieves higher
raw retrieval effectiveness, so SSR is not presented as a universally
superior lexical retriever. In end-to-end evaluation, SSR and BM25
candidate pre-selection show comparable observed effectiveness under the
same maximum candidate budget $K=20$; no statistically detectable
difference in end-to-end correctness is observed after Holm correction.
At $N=5000$, both candidate-pre-selection pipelines remain operational,
whereas none of the 200 full-catalog LLM requests completes
successfully under the evaluated model/API configuration.

These results show that the main candidate-context reduction arises from
effective candidate pre-selection rather than from SSR scoring alone.
The contribution of this work is the formulation and systems-level
characterization of candidate-context scalability for Agent Card
discovery, together with SSR as an Agent-Card-native, deterministic,
training-free, field-derived, and score-traceable retrieval mechanism.
The conclusions remain limited by the synthetic nature of
SSR-AgentBench, the document-derived Agent Card abstraction used with
MS MARCO, retrospective validation of the fixed candidate budget, a
single LLM/API configuration, and the exclusion of offline index
construction from online latency. Future work should evaluate native
Agent Card collections, prospective budget selection, query-adaptive
and hybrid retrieval, and dynamic or distributed registries.

\bibliographystyle{IEEEtran}
\bibliography{referencias}

\end{document}